# Relaxation of potential, flows, and density in the edge plasma of CASTOR tokamak

M. Hron[1], V. Weinzettl[1], E. Dufkova[1], C. Hidalgo[2], I. Duran[1], and J. Stoeckel[1]

[1] *Institute of Plasma Physics, Association Euratom / IPP.CR, Prague, Czech Republic*
[2] *Euratom Ciemat, Madrid, Spain*

Decay times of plasma flows and plasma profiles have been measured after a sudden biasing switch-off in experiments on the CASTOR tokamak. A biased electrode has been used to polarize the edge plasma. The edge plasma potential and flows have been characterized by means of Langmuir and Mach probes, the radiation was measured using an array of bolometers. Potential profiles and poloidal flows can be well fitted by an exponential decay time in the range of 10 - 30 µs when the electrode biasing is turn off in the CASTOR tokamak. The radiation shows a slower time scale (about 1 ms), which is linked to the evolution in the plasma density and paticle confinement.

**Introduction**

The mechanisms that control the generation of the radial electric fields ($E_r$) and damping of the $E_r \times B$ sheared flows represent a key issue for understanding the transition to improved confinement regimes in magnetically confined plasmas. There has been an intense discussion on the different driving and damping mechanisms of radial electric fields and plasma flows in the plasma boundary region of fusion devices. Both neoclassical (e.g.. ion orbit losses[1]) and anomalous mechanisms (i.e. anomalous Stringer spin-up[2,3], Reynolds stress[4,5]) have been considered as candidates able to explain the generation of sheared flows. Atomic physics via charge-exchange momentum losses[6,7], parallel viscosity (magnetic pumping)[8], and turbulent viscosity could explain the flow damping physics.

This paper reports the measurements of the relaxation times of plasma profiles after biasing turn-off in the plasma edge region of the CASTOR tokamak. Two biasing schemes have been used to modify the edge plasma parameters and confinement, first, with electrode inside the plasma comumn, i.e. inside the separatrix and next, with electrode placed in the vicinity of the separatrix (separatrix biasing).

**Polarization experiments**

In the CASTOR tokamak (R = 0.4 m, a = 85 mm, $B_t$ = 1 T, $I_p \approx$ 9 kA, $q_a \approx$ 10), the poloidal rotation and plasma potential have been modified by means of electrode biasing in the plasma edge region[9]. The electrode, located either in the vicinity of the last closed flux surface, at r = 65 – 75 mm or in the confinement region at r = 45 – 60 mm, was biased ($U_B$ = -400 ÷ 250 V) with respect to the vacuum vessel. Typical radial currents drawn by the electrode were up to about 30-40 A. Edge density and electron temperature were in the range $n_e \approx 1 \times 10^{12}$ cm$^{-3}$ and $T_e \approx$ 10-20 eV.

**Diagnostics**

Langmuir probes are used for the presented potential and flow measurements. The floating potential profile was measured using a rake probe with 16 single Langmuir probe tips that are radially separated by 2.54 mm, covering both the SOL and edge plasma regions (r = 65 – 95 mm). Further, a Mach probe of Gundestrup type was used for ion flows measurements[10]. The ion collecting surface of this probe is a nearly continuous cylindrical conductor (a copper tube of diameter 11.7 mm) divided into eight segments separated by 0.2 mm gaps. These segments are biased negatively to measure the ion saturation current.

A one-dimensional array of AXUV diods, acting as bolometers, was used to measure the radiation of the plasma column. A pin hole camera with 16 bolometers views the whole plasma cross-section from the bottom of the torus. The spatial resolution at the equatorial plane is 12 mm, signals are integrated along the line of sight.

The temporal resolution of all these diagnostics is 1 μs.

**Relaxation phenomena – floating and plasma potentials, electric fileds**

The relaxation times of floating potential and plasma rotation have been investigated after a sudden electrode biasing turn-off[11].
The maximum change in the floating signals $U_{fl}$ with biasing appears in the proximity of the polarization electrode and its perturbation extends radially approximately 1 cm both inward and outward the plasma column. Here, the radial electric field ($E_r$) has been computed using the expression:

$$E_r = -(\nabla U_{fl} + \alpha \nabla T_e)$$

where $U_{fl}$ is the floating potential, $T_e$ is the electron temperature measured by a swept Langmuir probe and $\alpha$ is a constant that depends on the probe material, on the ion species and its temperature and the secondary emission coefficient[12]. Fig. 1 shows the decay of the floating signals and of the radial electric field as measured by the rake probe. As the radial profiles of the floating potential and the radial electric field are modified mostly in the vicinity of the electrode position[13] (r = 75 mm), only the most influenced probe signals are shown.

The time evolution of the floating potential decay can be fitted to a function with the following shape:

$$U_{fl}(t) = U_{fl}^B \exp(-time/\tau) + U_{fl}^{OH}$$

where $U_{fl}^B$ and $U_{fl}^{OH}$ represent the mean floating potential values at the end of the biasing phase and after it, i.e. back in the ohmic regime, respectively, $\tau$ is the exponential relaxation time (Fig. 2). This fitting procedure has been done for floating potential signals measured at several radial positions in proximity of the polarization electrode. As apparent from Fig. 2, the experimental results show that the floating potential signals as well as the radial electric fields follow very well an exponential decay with a characteristic time in the order of 10-20 μs.

**Relaxation of plasma flows**

The ion mass flow was measured by the standard arrangement of the Mach probe of the Gundestrup type[14] (Fig. 3). The perpendicular Mach number ($M_\perp$), i.e. the perpendicular flow, is related to the ratio of ion saturation currents measured by the probe segments facing the poloidal direction. The parallel Mach number ($M_{||}$) is related to the ratio of the upstream and downstream segments.

The experiments have shown that flows at the end of the polarization period inside and outside the separatrix behave in a different way. The poloidal flow velocity decreases in the SOL (r = 84 mm), while it shows an increase in the edge plasma (r = 71 mm). As the poloidal flow in the CASTOR tokamak is dominated by the $E_r \times B$ drift the sign of change of ion current ratio (change of flows) is consistent with the change in $E_r$ at the two radial positions. The measured relaxation times of the poloidal flows are in the range of 10 – 30 µs. The parallel flow velocity remains unchanged in the SOL and shows a slight increase inside the separatrix.

**Relaxation of the enhanced density and in the radiation**

The density is enhanced by a factor of about 1.5 during the biasing period. After the termination of the biasing, the density starts to decay with a time scale related to the particle confinement (see Fig. 4). The density returns to the ohmic level during approximately 1÷2 ms. The radiation observed by the array of bolometers decays in a similar time scale: Fig. 5 shows in a spatial-temporal plot the change in radiation due to the biasing and the subsequent decay of the increase. The decay is fited using an equivalent formula like for the floating potential decay in the probe measurements. The obtained characteristic time is in the range of 0.1 ÷ 5ms, showing faster relaxation for higher biasing voltages (both positive and negative).

**Discussion and conclusions**

It is important to compare the measured relaxation times of the floating potential and the plasma flows with the characteristic time of the plasma turbulence, the parallel viscosity, and the charge exchange mechanisms. The characteristic time scale of the viscosity would depend on the plasma conditions (e.g. collision time and safety factor) in case of the magnetic pumping[14] whereas the eddy viscosity is expected to be related with the time scale of the energy transfer between different turbulent scales (i.e. a few turbulence correlation times) and the atomic physics damping rate is directly related with the neutral density[15].

The correlation time of plasma turbulence (floating potential signals), defined as the full width of the autocorrelation fluctuation on its half maximum, is in the range of 5 µs in the CASTOR plasma edge region. The momentum loss rate due to the atomic physics mechanisms (charge exchange) can be expressed as $\tau_{cx} = (<\sigma_{cx} v_i> n_n)^{-1}$, where $n_n$ is the neutral density. Assuming $T_e \approx T_i$ and $n_n \approx 10^{17}$ m$^{-3}$, it follows $\tau_{cx} \approx 500\text{-}1000$ µs. Poloidal flows can be damped by parallel viscosity (magnetic pumping) and several of the neoclassical viscous forces can be found in the literature[16,17]. In the collisional regime, the poloidal momentum damping time can be estimated as

$$\tau_\theta = (2 + \frac{1}{q^2}) \frac{1}{\omega_{T_i}^2 \tau_{ii}}$$

where $\omega_{T_i}$ is the ion transit time and $\tau_{ii}$ is the ion collision time[8]. For CASTOR edge plasma parameters, $\tau_\theta$ is in the order of 100 µs.

Present results show that the experimentally measured fast damping times of the floating potential, plasma potential, electric fields, and poloidal flows are in the same order of magnitude in the plasma edge of the CASTOR tokamak (10 – 30 µs) that is less than the expected damping times based on magnetic pumping mechanism (in the range of several 100 µs) and on atomic physics via charge exchange (in the range of 500 µs). On the other hand this is slightly larger than (or comparable to) the correlation time of plasma turbulence (5 µs).

This finding suggests the existence of anomalous (turbulent) mechanisms in the damping rate of radial electric fields and poloidal flows in the plasma boundary of fusion plasmas.

In addition to the fast relaxation time scale, a slow relaxation time scale (in the order of 1 ms) was observed on the plasma radiation. This slow scale is connected with the density evolution after the biasing switch-off.

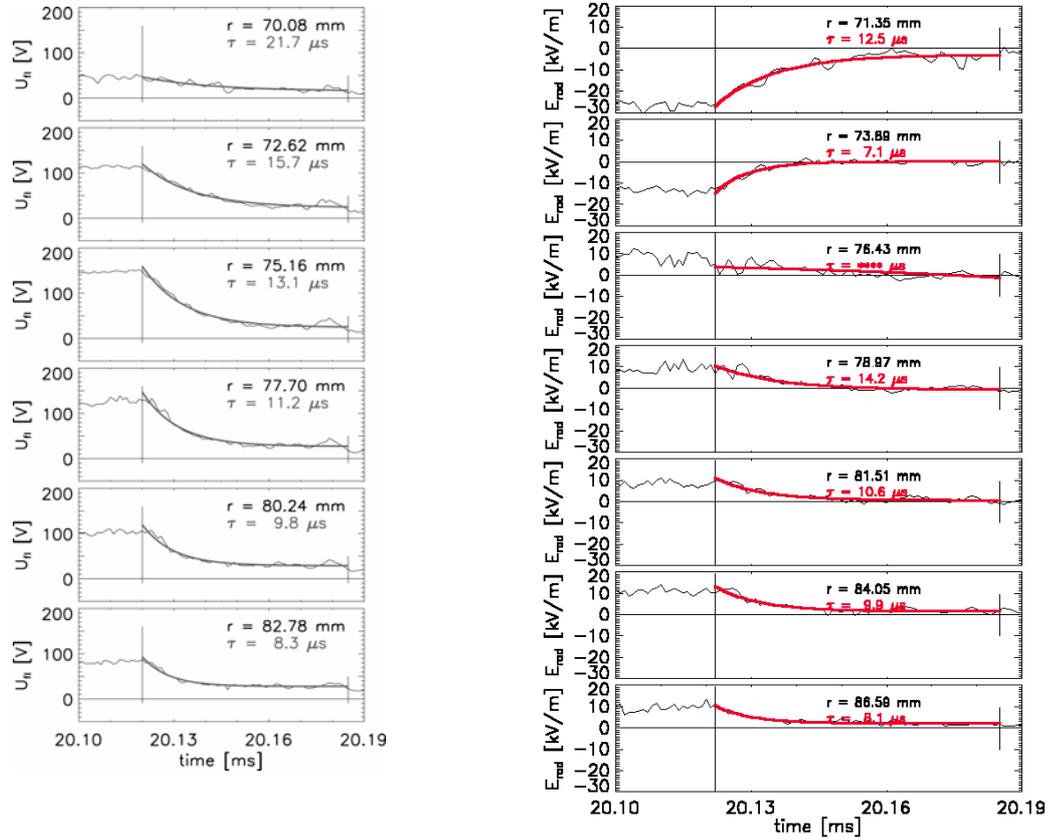

*Fig. 1: Floating potential (left) and radial electric field (right) relaxation after the biasing voltage switch-off measured by Langmuir probe tips of the radial rake probe on the CASTOR tokamak. From top to bottom – the tips are located insight the separatrix and outward.*

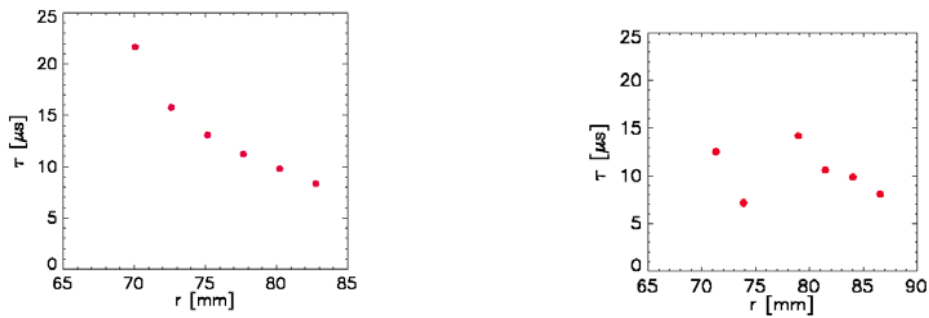

*Fig. 2: Characteristic relaxation time of floating potential (left) and radial electric field (right) decay after the biasing switch-off (CASTOR).*

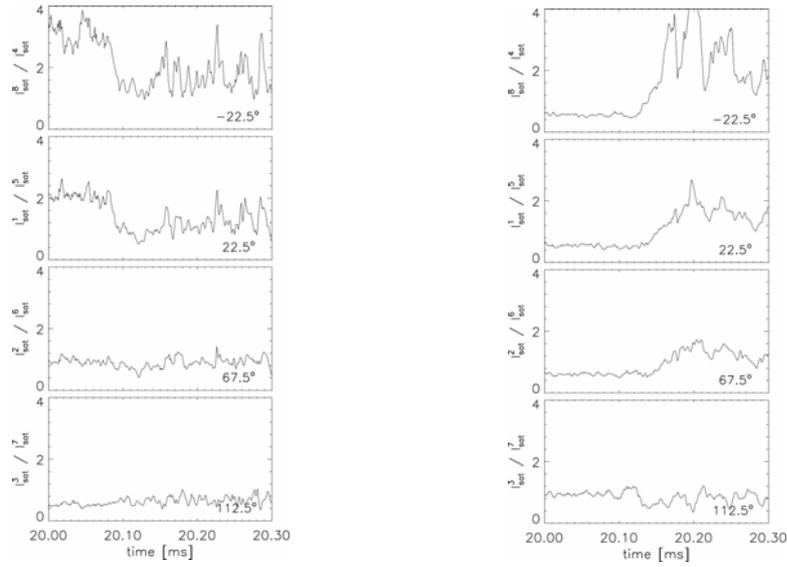

Fig. 3: *Flow measurement (evolution of the $I_{sat}$ current ratios) using the Gundestrup probe on the CASTOR tokamak: signals related to the perpendicular flow – $M_\perp$ (two top pannels) and to the parallel flow – $M_{||}$ (two bottom pannels). Data obtained in the SOL, r = 84 mm (left column) and inside the last close flux surface, r = 71 mm (right column).*

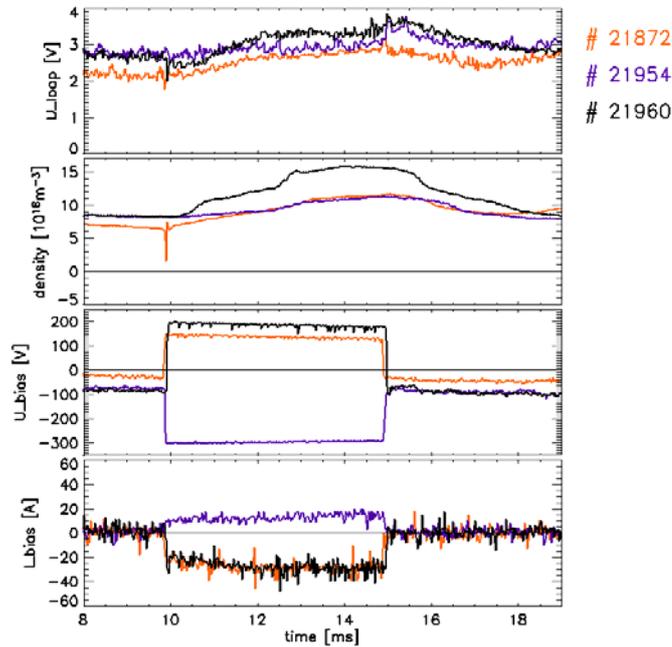

Fig. 4: *Evolution of main plasma parameters – loop voltage, line averaged density, biasing voltage, and biasing current for three shot. For two of them, the biasing voltage is positive (#21872, #21954), negative biasing is applied in one shot (#21960).*

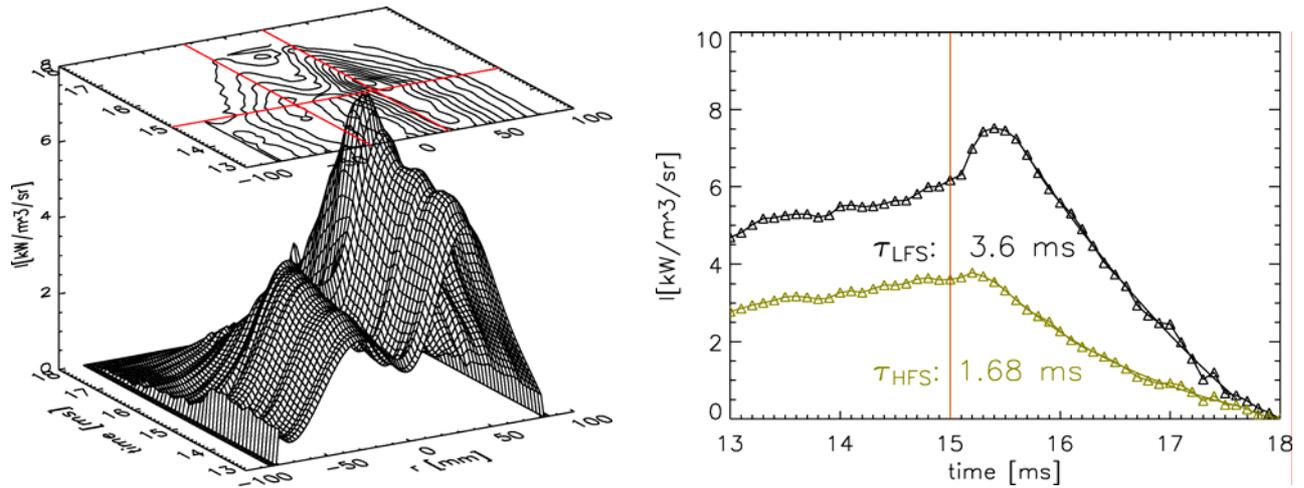

*Fig. 5: 3-dimensional plot of the plasma radiation during and after the biasing phase of the discharge (ohmic background is subtracted), #21872, $r_B=70mm$, $U_B=+140V$ (left). Decay of the intesity at high and low field sides of the torus (right).*

**Acknowledgement:**

This work was supported by the Grant Agency of the Czech Republic, contract number 202/03/0786.